**Manoj Devare**
Doctor of Philosophy, Professor, Head of Institute, Amity Institute of Information Technology,
Amity University Maharashtra, Mumbai, India.
ORCID ID 0000-0002-9530-3914
*mhdevare@mum.amity.edu*

# AMALGAMATION OF INDIRECT GAMIFICATION INTO DEVELOPMENT AND OPERATIONS (DEVOPS) COURSE TEACHING

**Abstract.** The work investigates the effect of indirect Gamification in Development and Operations (DevOps) Course teaching. The software development team consisting of multiple contributors coordinates in the collaborative work environment to achieve specific predefined goals in a regulated and controlled fashion. In DevOps Course, the Installation and configuration of the various software tools are useful in the context of application development. To teach DevOps Tools precisely to achieve the learning outcome is a skilled task for a trainer. From the perspective of the learner, the learning environment has to be encouraging and exciting. Pedagogical techniques and ICT tools play an essential role during the knowledge transfer process. The motivation and reward to the Student are essential in achieving the learning outcomes of the Course. The author used the third-party Competition as an indirect Gamification technique to achieve the learning outcomes of the Course. The author encouraged the students to participate in the Hacktoberfest Competition to use the practical skills learned in the Course. The Under Graduate (UG) and Post Graduate (PG) Students have gone through the regular sessions of DevOps. For the participation purpose, both groups of students communicated on one platform. The students who succeeded faster during the GitHub Pull Request (PR) submission shared their experiences with other participants. The PG students participated 68.75% higher than UG students. Minimum of four PR submissions on GitHub and acceptance by the repository maintainer are the task completion criteria. The active participation of the small number of UG students became a motivational factor for the PG students. The Gaussian distribution on the marks obtained by the experimental group shows the absence of outliers. The research shows that the effectiveness of indirect Gamification depends on the age group, level, course content, and learning environment. The participation of a faculty member in the Competition during the learning activity boosts the desire of the Student to complete the task. The experimental group of 15 Students has outperformed in terms of the marks obtained compared to the control group of 52 students.

## 1. INTRODUCTION

Gamification is an activity of adding game mechanisms into services or products to motivate or engage the users for positive outcomes. Gamification improves the natural desires of the participants. Gamification has effectively established significance in various fields such as Human resources, Finance, Banking, Marketing, E-Commerce, Healthcare, Personal Growth, Societal Improvement, and Pedagogy. The critical game mechanics give badges, challenges, leader boards, points, levels, and rewards to encourage participation. This work is specific to applying the indirect Gamification in the highly technical Computer Science (CS) and Information Technology (IT) course on DevOps.

The DevOps, Agile Software Engineering Methods, Tools, and Techniques are essential skills for Software Developers, and IT teams working in the Operations domain. The DevOps tools Git, GitHub, Maven, TestNG, Eclipse, and Jenkins help automate Software Engineering processes. The students are actively engaged in the DevOps Course with theory concepts, demonstrations, and hands-on practice. The teaching of the CS and IT courses during the COVID-19 lockdown is carried out using online meeting tools and supporting technologies like Stream to store the video lectures in the Cloud.The teaching and learning during the immobility was a didactic experience for students and faculty members. The data availability upto 1.5 GB to 2 GB was essential, and conducting all sessions listed in the timetable was a hectic task for teachers and students.

**The problem statement.** The learning outcomes of the highly technical courses taught in UG and PG Programs should be achieved effectively.

The current pedagogy frequently results in an uninteresting classroom environment in the case of highly technical courses where various software has to be learned. If the problem is ignored, the Student may miss the targeted learning outcomes. The quality of teaching may suffer, as a result.

Finding the combined effect of motivation, reward, and Indirect Gamification in the teaching and learning of technical courses like DevOps is investigated in this work. The effect of amalgamation of the Indirect Gamification into teaching and learning by introducing the Hacktoberfest competition is the primary intention of this research work. This proposed research aspires to explore the option of Indirect Gamification in pedagogy. The faculty trainer does the encouragement, motivation, and monitoring during the course tenure. The marks obtained by the experimental group and control group are used to interpret the effectiveness of the technique applied in the pedagogy.

**Hypothesis.** The use of Indirect Gamification in the pedagogy of highly technical DevOps courses in CS and IT improves the academic performance of the Student.

**Analysis of recent studies and publications.** Gamification is used as a technique by the Trainers and Researchers at different levels of teaching in CS and IT courses. An extensive review on the use of Gamification in Science education and education management is available. Kalogiannakiset al.[1] carried out a the review of 177 papers on the Gamification used in Science Education. Da-Silvaet al.[2] have done a literature review of 241 papers on education management. Dichev and Dicheva [3] have done a critical review of 88 papers regarding the status of Gamification in education. Trinidad et al. [4] have done a bibliometric study to describe the scientific and structured evolution of Gamification.

The researchers have tried Gamification during the teaching of the CS and IT Courses, such as teaching C-Language programming, Artificial Intelligence, and Cyber Security courses. Ibáñez et al.[5] found the moderate positive impact of using the Gamification in learning of the C-Programming Language. Chubarkovaet al.[6] have tried Gamification in AI Education. Roset al.[7] tried the Non-compulsory Gamification for the effective teaching of the Cyber Security course. Oktaviati[8] found that implementing gamification elements in learning can motivate and engage the Student learning CS.

The effect of the Gamification and its relation with the Gender, age groups, and levels is reported. Khan et al.[9] reported Digital Game-Based Learning and Gamification in teaching and learning. The results show that the Game-Based Learning application did not equally work for all students. The girls outperformed boys in engagement and learning outcomes. The study gives insights into the development of better educational games to promote student learning. Connolly et al.[10] reported the results of a systematic literature review on the impact of gamified activities on young learners. Lam et al.[11] conducted an experiment involving four secondary students to check the effectiveness of the digital game. After seven weeks of observation, it was found that digital games improve the online contribution and writing performance of the Student. The experiment was done based on the blended learning and gamification approach. The significant improvement results are found in the writing skills for the population size of 30. Hasanet al.[12] use online discussion environment of the Gamification and Moodle Learning Management System (LMS) during the term paper evaluation of the PG students.

Darejeh[13] used Gamification in online education. Hasanet al.[14] discussed that in the collaborative work environment, feedback and Gamification are essential. Auvinen et al.[15] tried giving achievement badges as a form of Gamification to reward and encourage the students. Cuevas-Martínez et al.[16] conducted a gamification experiment over four years in Spain. The use of Gamification during the student assessment and video games like Super Mario positively affects the students' dropout ratio. García-Iruela et al.[17] used Gamification for creating the research interest among the students. Pacheco-Velazquez[18] developed the challenging game as a pedagogical design. The experiment shows that the game design in the logistic area promotes the Student's self-directed learning skills. Romero-Rodríguez et al.[19] used Gamification strategies to achieve better effects from Massive Open Online Course usage in teaching.

Adrian et al.[20] developed an esthetically pleasing user interface of the gamification tool at the US Air Force Academy. According to [21],[22] the learning outcomes and student satisfaction play an essential role in teaching the challenging Course in Computer Science. The simplified user interface, the Client-Server suite, and dubbed Live Classroom simplified the everyday tasks performed by students and teachers. Villagrasa[23] has worked on the students' collaborative work, virtual teacher environments, and visualization 3D on the webGL for effective teaching. Vranešićet al.[24] & Varannaiet al. [25] used the Kahoot system for the UG programs in Information and Communication Technology (ICT) and awarded extra marks to the students based on their performance.

The investigation of the literature shows that the effect of third-party Competition in the course learning outcomes is not reported. The Trainer has brought the UG & PG students to one platform and used the participation of the UG student as the influencing factor to encourage the PG students. The kind of research work is not reported either. The proposed research of amalgamation of the Indirect Gamification into the DevOps Teaching unique work adds new knowledge to the existing pedagogy.

**The Research Goal.** This work aims to investigate learning outcomes as an effect of amalgamation of the indirect Gamification through the third party competition in the DevOps Concept course at the UG and PG level.

## 2. THE THEORETICAL BACKGROUNDS

Effective knowledge transfer techniques are seldom found in the CS and IT teaching and learning processes. There is a general assumption that Students and the Faculty of these fields are more aware of software tools, hence they use suitable tools and techniques effectively. However, the learning abilities of different students are not similar. Therefore, well-tested techniques and methodologies are required to be employed. Technically, students and faculty trainers working in this field expect some other methodologies to learn and teach the content of the courses in a better manner. In addition to the traditional pedagogical methodologies, Gamification can help both students and teachers.

The hype cycle, such as one by Gartner, gives intuition of the recent trending technological advancement topics in the product and service industry. The inclusion of these topics in the curriculum is the right step to develop the human resource, as per the industry's requirement. The Job Description and Interaction with the IT Industry Leaders are the source of information to understand the skill required in the software Industry. These three sources and market research help in the Curriculum design of the UG and PG Programs. The course content of DevOps Concepts is designed by considering these requirements of the IT industry.Considering this scenario, tools such as Git, GitHub, Maven, TestNG, Eclipse and Jenkins are part of the DevOps Concepts Curriculum. The inclusion of the DevOps tools as a part of regular teaching in the CS and IT courses will help the students while working in the IT industry when they join as freshers. The Course is a part of the UG and PG curriculum to boost confidence among the students. Because the contribution in terms of the PR submissions on the version controlling system is also related to the performance of the employee, it is one of the mechanisms in the IT industry to count the contribution by the Software Developer.

The DevOps Concepts is an advanced level Course that needs knowledge of the Programming language, Open Source Software, and Software Testing Methodologies. Moreover, it requires the skills to integrate all the tools used in the Agile and Traditional Software Development Methodologies in Continuous Integration (CI) and Continuous Development (CD) fashion. The Course is designed as the two credits for the theory and one credit for Laboratory work. The faculty trainer covered the theoretical concepts in two hours of teaching. The faculty trainer also engaged the students for the two hours of practical demonstrations and monitoring hands-on practice. The Course is taught to the PG students of Master of Computer Applications (MCA) and UG Students of Bachelor of Computer Applications (BCA). The DevOps course is in the first semester of the two-year PG program and in the third semester of the three-year UG program. All the UG and PG students had a specialty in Computer Science and IT.

The knowledge of Software Development Life Cycles (SDLCs) is the foundation block for Software Engineering. The practice of using SDLCs is a traditional method used in the Software Industry. Overtime, the processes evolved, and agile methodologies are a regular practice in the software industry. After discussions with the industry leaders and the practitioners, it is found that "Scrum" is the important technique followed.

The job roles such as Product Owner and Scrum Master are essential skills for the IT Team. The recent developments in the field of DevOps and the integrations of the automation tools are used at the various stages of product development. The Product Owner and the Scrum Master drive the Development Teams by floating the Functional and non-functional requirements.

## 3. RESEARCH METHODS

The online conduction of the DevOps Concept course happened between18 August 2020, and 24 December 2020. The Hacktoberfest 2020 competition came precisely in the middle of the course tenure. The author has carried out the research work at the Information Technology School of the University where the author is currently working. The end-of-semester examinations were conducted in January 2021. The DevOps Course is a part of the UG and PG programs at the IT school of the University. The effectiveness of a technique is measured in terms of the successful participation and completion of the task by the Student. Moreover, the outcome was also measured in terms of the marks obtained by the Student during the continuous evaluation and final results declared after the end-of-semester examination.

For the experimental purpose, the students of the BCA-UG and MCA-PG programs were selected. Initially, the encouragement to participate in the Competition was given to all the students of UG and PG. However, more participation was observed from the PG Students compared to UG students. The sample was divided into the experimental group and control group, as shown in table 1.

*Table 1*

**Students Count as Experimental Group and Control Group.**

| Level | Total Students | Experimental Group | Control Group |
|---|---|---|---|
| UG | 50 | 02 | 48 |
| PG | 17 | 13 | 4 |

The students who participated in the Competition were in the experimental group. As shown in figure 1, there were 02 students from UG and 13 students from PG who participated in the Competition. The students who did not participate in the Competition were in the control group. There were 48 students from UG and four students from PG in the control group.

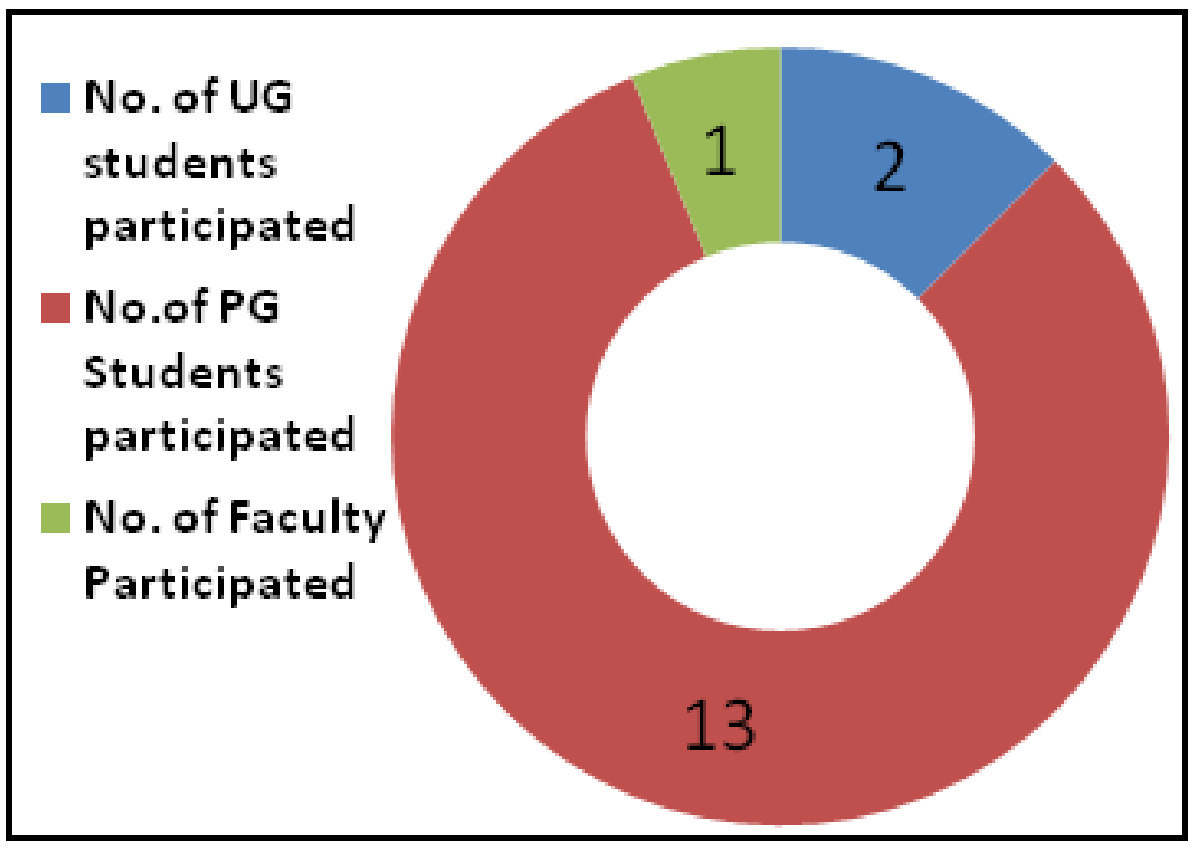

*Figure 1. Hacktoberfest Participant Count Details*

Figure 2 shows that out of the total participating PG students, there were 5 girls (38%), and 8 boys (62%). Participation is not simply a skill to work in the collaboration – it also polished knowledge of the students' programming language and involved quality program writing practice.

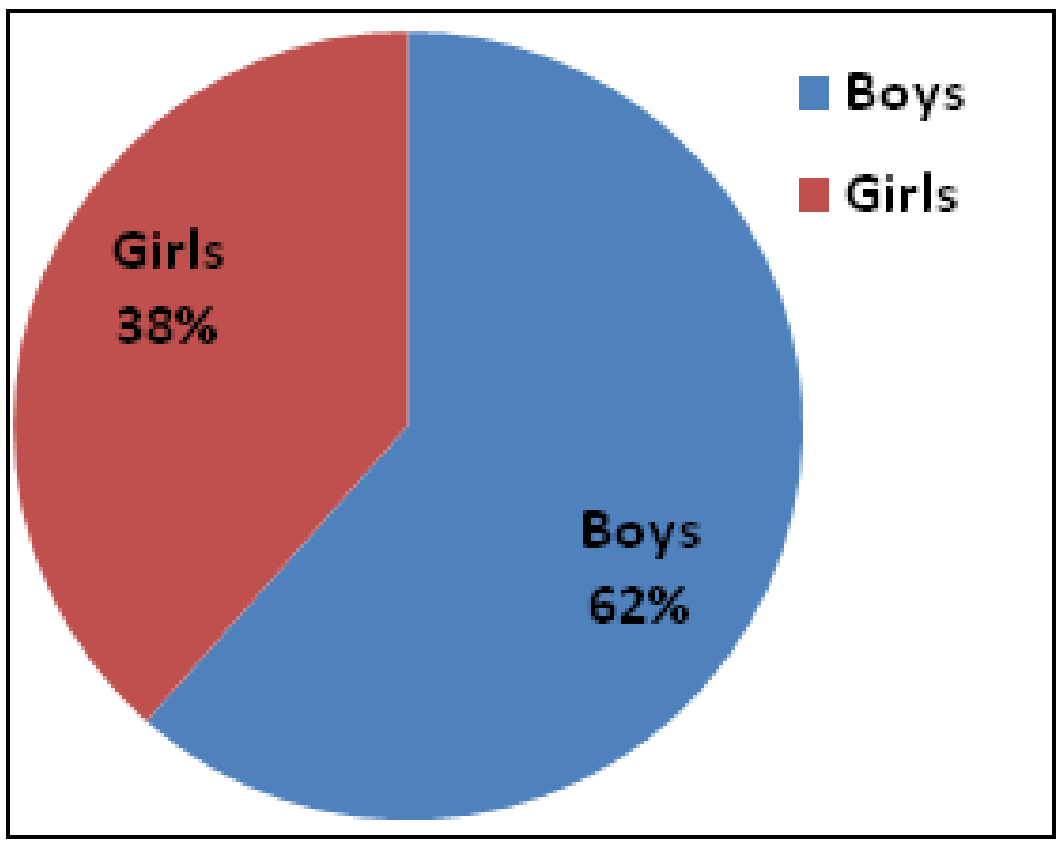

*Figure 2. Girls and Boys Ratio in Hacktoberfest participation*

### 3.1 Teaching-Learning process of the DevOps Course

The Trainer of the DevOps course requires the prerequisite knowledge of the Open Source projects working under different Licensing policies. Faculty trainers need to have prerequisites of Software Engineering and Software Project Management. Eventually, the Trainer can cover the topics on CI/CD Tools and Techniques. The Course starts with the basics; however, the practical work also goes hand-in-hand. The Installation of the various Open Source DevOps Tools is an essential skill. The successful Installation of these tools requires troubleshooting skills by looking into the online postings.

### 3.2 Important DevOps Tools

There are several open-source and proprietary software tools available for version control. However, Git and Github are more popular among the programming community. There are two approaches to use Git and GitHub, namely, using, from GitHub side or start using from the Git Tool installed on the computer of the user. Github is the Distributed as well as central repository for the collaborative project work. The critical learning outcomes for Git and Github are Push, Pull, forking, cloning, creating log records, merging, and branching operations.

Maven is the build and project management tool most suitable for Java developers. The dependencies on the third-party Java Archives files can be quickly resolved by installing the Maven. Maven can be configured as Project Object Model and integrated with the Integrated Development Environment tools like Eclipse. The TestNG testing methodologies, parallel execution of the testing threads, and annotations are part of the Course. The configuration of the Java projects for the Local, Central, Remote repository, build the project, deploy as per the target are the critical learning outcomes for Maven Learning. The TestNG is the tool installed in the Eclipse with the Maven projects.

Jenkins is the automation tool used for the configuration of the various Jobs to be triggered as the result of the updates in the development events, such as remotely executing the Job, Windows batch file execution, execution of the jobs after the specified time interval, executing the Job after a particular Job, triggering the build after linked updates found on the GitHub, and Sending the notification as an email. These CI/CD integration methods of the automation pipeline are essential for the smooth delivery of the updates in software development.

### 3.3 ICT Tools used in teaching

The ICT Tools such as MS-Teams, LMS, Powerpoint presentations are helpful for teaching. Whatsapp is used for coordination and communication among students and faculty members. The assignments are submitted on the LMS developed in-house. The deadlines-based assignments are evaluated on Moodle LMS, as shown in figure 3.

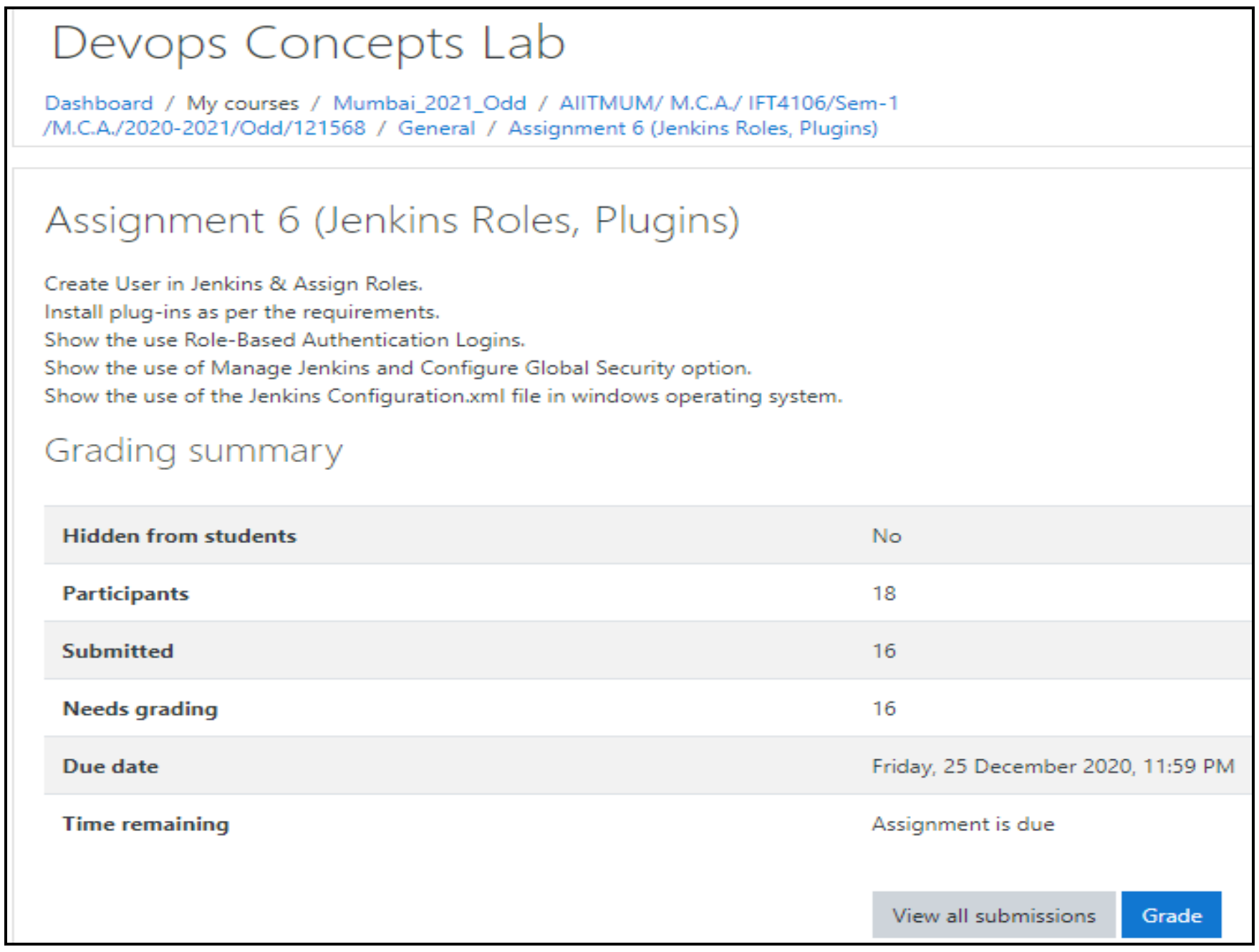

*Figure 3. Assignment Submission records on LMS for DevOps Course*

### 3.4 Indirect Gamification

The author used Indirect Gamification by introducing Hacktoberfest 2020 Competition as a part of the learning outcome of the Course. The direct reward points are not offered to the learner, but the prizes offered by the competition organizers helped to provide the rewards. The outcomes are measured in terms of successful PR submission to GitHub. The Course learning outcomes are made more attractive to develop the skills among the students, such as Git and Github tools for the version controlling. Peer learning has worked significantly when few students lag in reaching the target of four PR submissions. The requirement of the quality PR submission is fulfilled by creating the programming code and unit testing by the contributor.

The Direct rewards are given by the organizers of the Hacktoberfest, such as T-shirts and Stickers from the sponsors. After few months, the postal packets of the prizes come as the home delivery. The activity of the PR submission process continues in October. The website of the Hacktoberfest shows the progress and maturity of the PR submission after the specified time interval.

### 3.5 About Hacktoberfest Competition

DigitalOcean launches Hacktoberfest competition. Participants in the Hacktoberfest represented 151 countries having thousands of unique skill sets. The Hacktoberfest competition is regarding the awareness among the software development community to use the version controlling system in the teamwork. The collaborative work in the team requires the central and distributed repository in the organization, such as Github. The participant has to send the four PRs in October in any time zone. The PR can be to any public repository on GitHub, not just the ones highlighted. The PR must contain commits made by the participant. If a maintainer of the repository does not report PR as spam, it is counted toward participation in Hacktoberfest.The quality of the code contributed by the participant has to be maintained by writing a neat and clean code.

In this case, the participant students and faculty members of the DevOps Course contributed to the Open-Source GitHub Public repositories by writing the programs in Python, Java, C++, and JavaScript. Therefore, the PR contribution, merging, and acceptance by the maintainers of the public repositories are the minimum requirement. However, the enthusiastic participants who are part of the population of this research work have contributed more than sufficient requirements of the PR submission.

### 3.6 Faculty Involvement at Every Stage

The indirect Gamification in teaching requires the faculty involvement to monitor progress and solutions provided by the students. Giving the proper direction, rectifying the errors made by the students is handled by the Trainer. The students also learn from other students in the batch. Few students can understand the open-source tools very quickly. The understanding of such a student assists other students in the class. Accompanying such a student with some student who is facing difficulty can help in a positive direction.The Faculty has an essential role in seeing whether the students are synchronized with other participants. The faculty trainer's role is to take the feedback from all such parties involved in the communication.

Whatsapp communication chat between the students and Faculty is shown in figure 4. The figure shows the feedback and progress monitoring of the participant students. Few students have faced technical difficulty during the completion of the task. The active students and Faculty have helped them to resolve the issues. A participant’s selecting the wrong labels may lead to disappointment, and the repository owner may not accept the PR. The selection of the active repository on Github is an essential criterion. If a participant selects the wrong repository, such as the maintainer is not looking into the PRs submitted by the participants for the longer duration, then all efforts go to waste. If the maintainer takes a longer time, then the competition duration finishes, and the participant experiences disappointment.

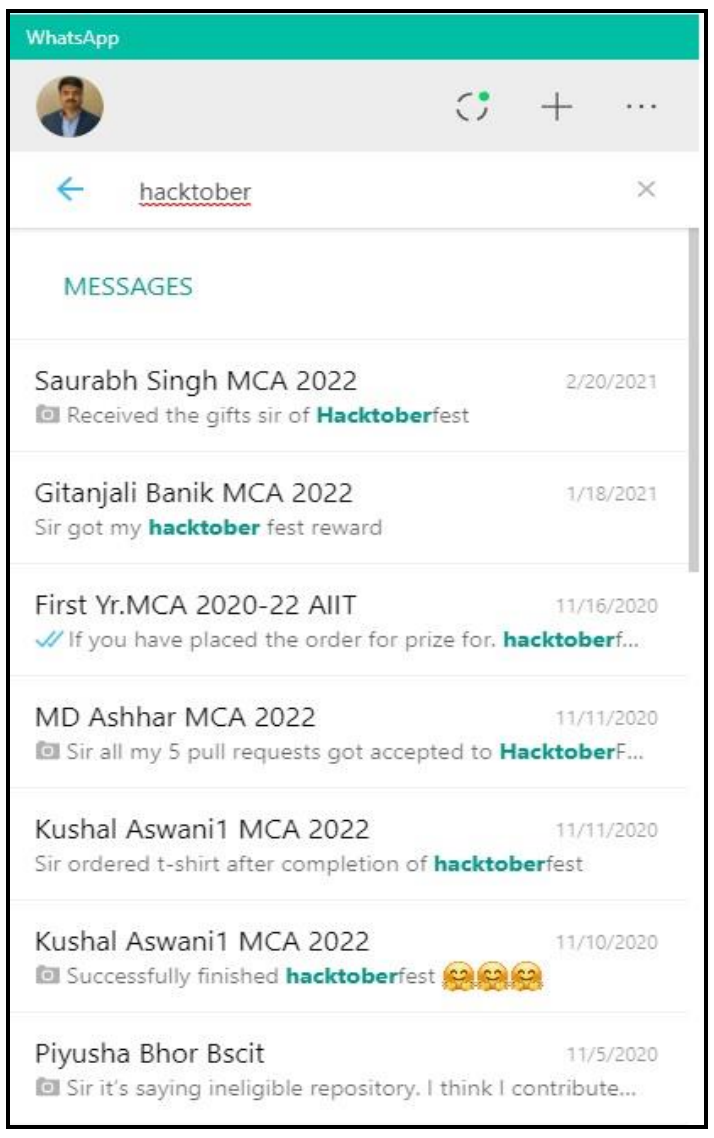

*Figure 4. Whatsapp chats with PG students for Hacktobefest Competition*

Here Faculty trainer participated along with the students to encourage and maintain the rhythm towards completing the task. Furthermore, the Faculty maintained the quality of the PRs by uploading the Java solution programs for the Graph Coloring of the Bipartite Graph Theorem, graph degree calculation using Handshaking Lemma, Direc's Theorem for detecting the Hamiltonian Circuit in a simple graph, a necessary condition for Eulerian Graph, and checking if the graph is Semi-Eulerian or not.The Faculty shared these contributions with all the participant students to make them aware that the Trainer chose high-quality Graph Theory and Combinatorial topics to participate in the Competition. Figure 5 shows the faculty trainer progress screen of the Hacktoberfest Competition.

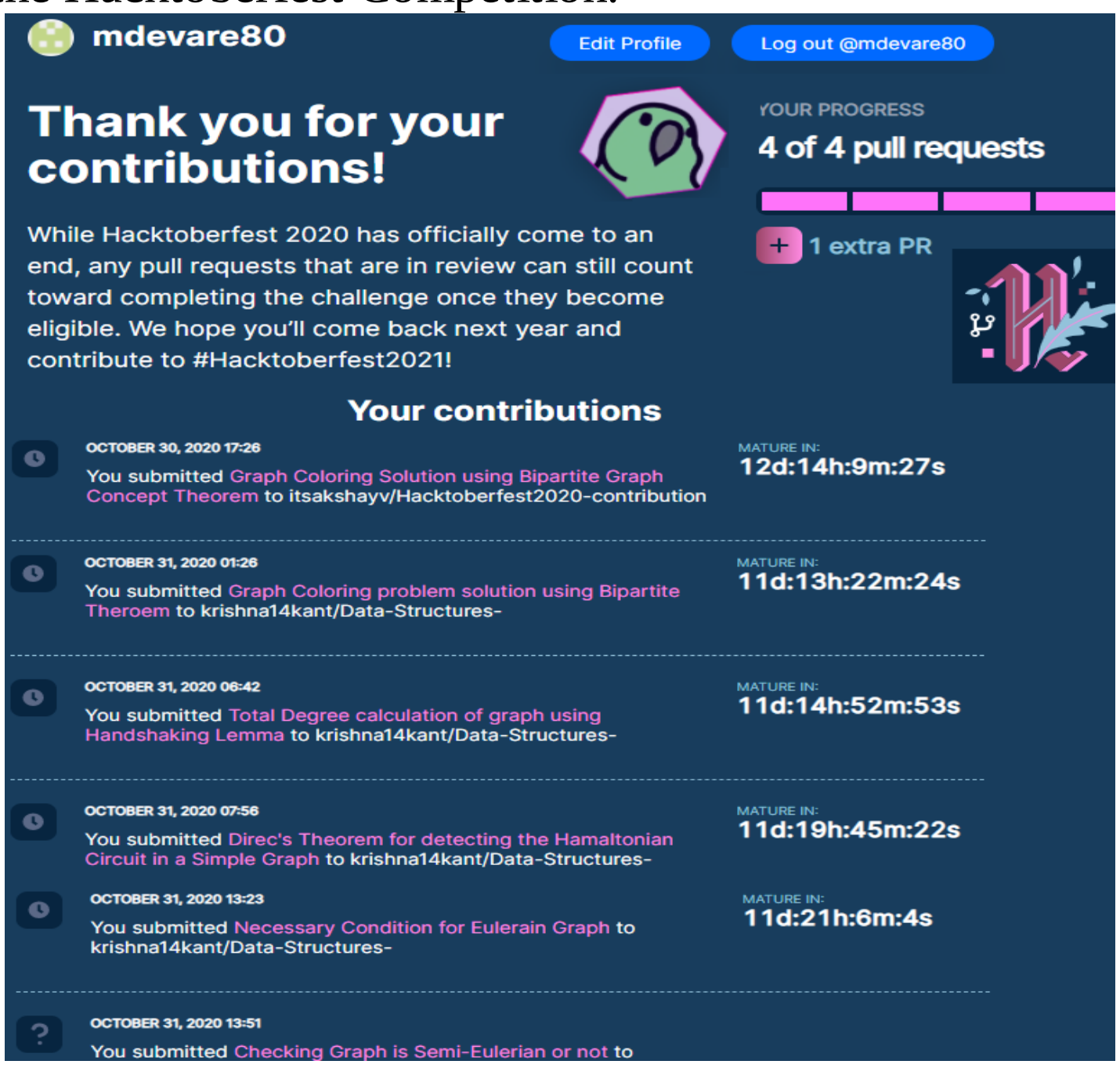

*Figure 5. Faculty trainer's participation in the Hacktoberfest PR submission on GitHub*

### 3.7 Motivation to the students

The motivation from the faculty trainer is an essential factor in maintaining the rhythm with the students. The trainer faculty used the words like "Well-Done," "Great Work," "Congratulations," "Good Work," "Keep Doing Good" to appreciate the students. The appreciation for the excellent work used in personal communication with the learner creates a rapport with the Student. Nowadays, faculty trainers may use Whatsapp, telegram, or similar tools for communication with the students. In addition, small images like "Claps" can be helpful to congratulate students on their achievement. Figure 6 shows the motivation provided by the faculty trainer during the Competition.

The most important part of the motivation is to announce the success of our students at the higher level of the organization. In addition, sharing the information of the achievement with participating and non-participating students can build a competitive environment.

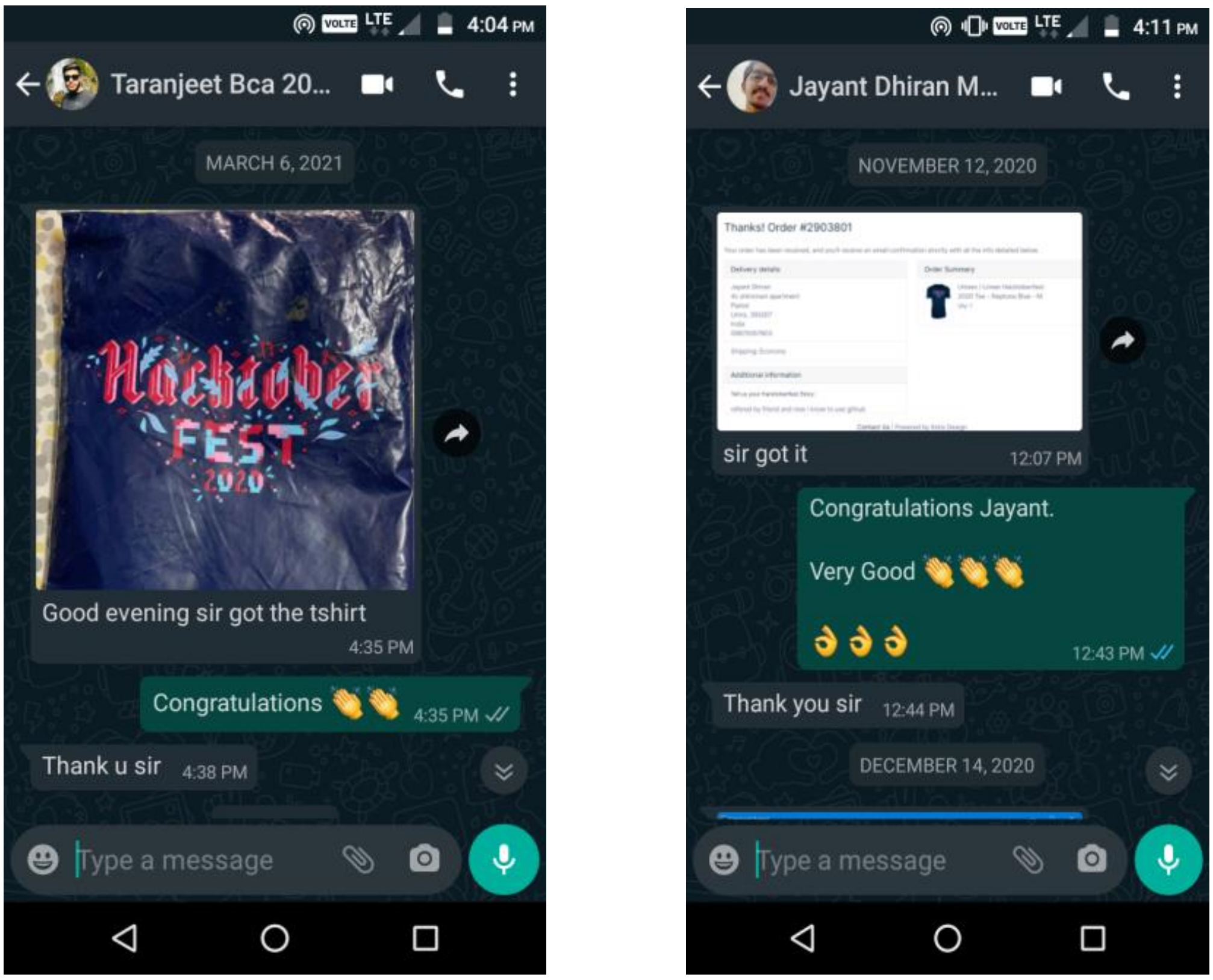

*Figure 6. Motivation and appreciation by the Faculty trainer to the students*

## 4. THE RESULTS AND DISCUSSION

The outcomes of the Course can be considered as tangible and intangible. The tangible outcomes are counted here as the Marks obtained and Prizes received by the UG and PG students. The effect of the motivation used by the Faculty is not calculated; however, the positive effects are observed by the Faculty. The intangible outcomes are important because they are related to the skill-based final placements of the students. Another outcome of the DevOps subject is excitement among the students to use the tools in post-course finishing and completing the end-of-semester examination. Thus, the experimental group well received the intangible and intangible outcomes.

**a) Marks Obtained as an Outcome**

The course examination evaluations are divided into continuous and end-of-semester examinations. A maximum of 30 marks is assigned for the continuous evaluations, whereas a maximum of 70 marks is assigned for the end-of-semester examinations. The continuous evaluation methodology used here is the multiple internal handwritten assignments, Final Journal Compilation of the practical work performed, end-of-semester examination for the practical work. The First outcome that we consider is in terms of the marks obtained by the students in continuous evaluations, end semester examinations, and overall. The Box and Whisker of the marks obtained by the experimental group out of 100 found that the median is at 81, and the third quartile is at 93, as shown in figure 8. The minimum marks are found at 72. As there is no outlier found on the lower tail and the upper tail, normally distributed data is found in the marks obtained.

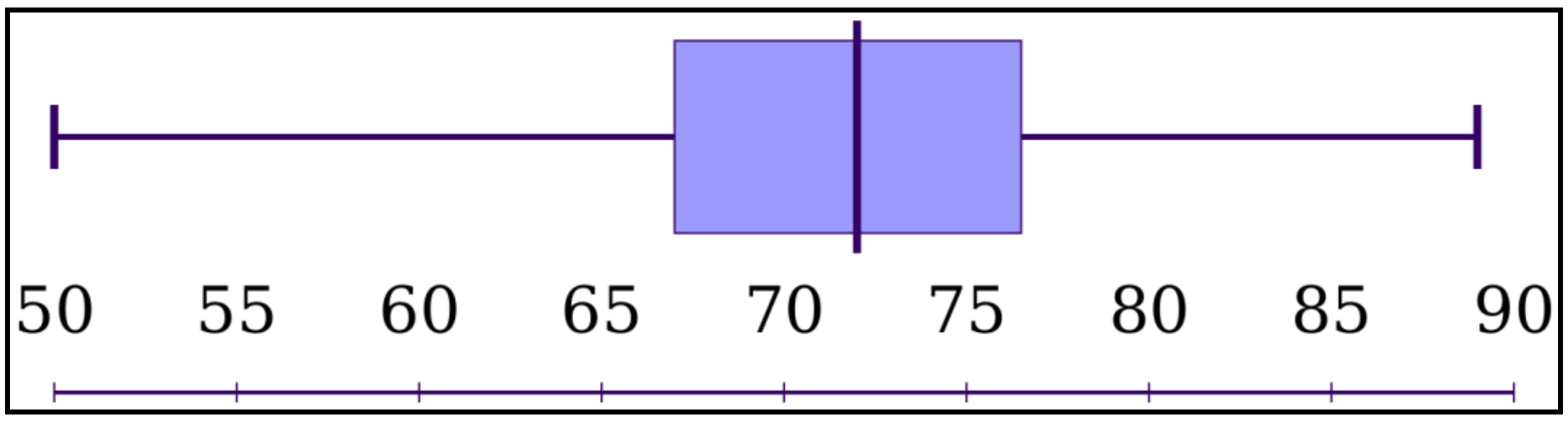

*Figure 7. Control Group marks out of 100 in DevOps Course*

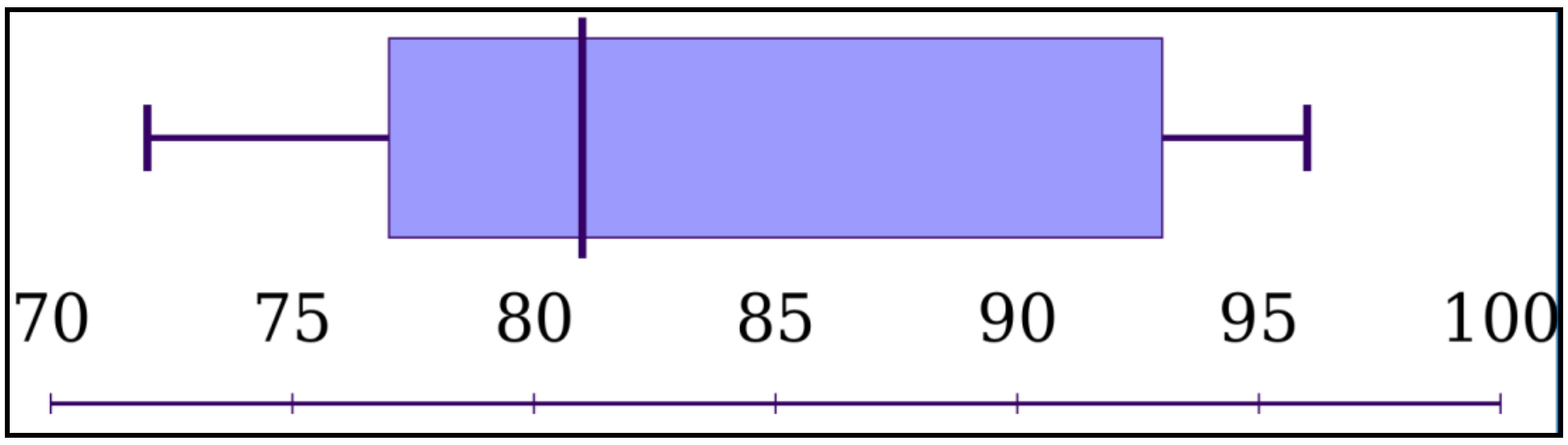

*Figure 8. Experimental Group marks out of 100in DevOps Course*

The Box-Whisker plot in Figure 7 & Figure 8 shows that the experimental group has more marks than the control group. Each quartile of the experimental group has more marks in overall performance. The distribution of the marks for the experimental group is skewed towards the right side. This indicates that the students who participated in the Competition are having more marks than the students who did not participate in the Competition.

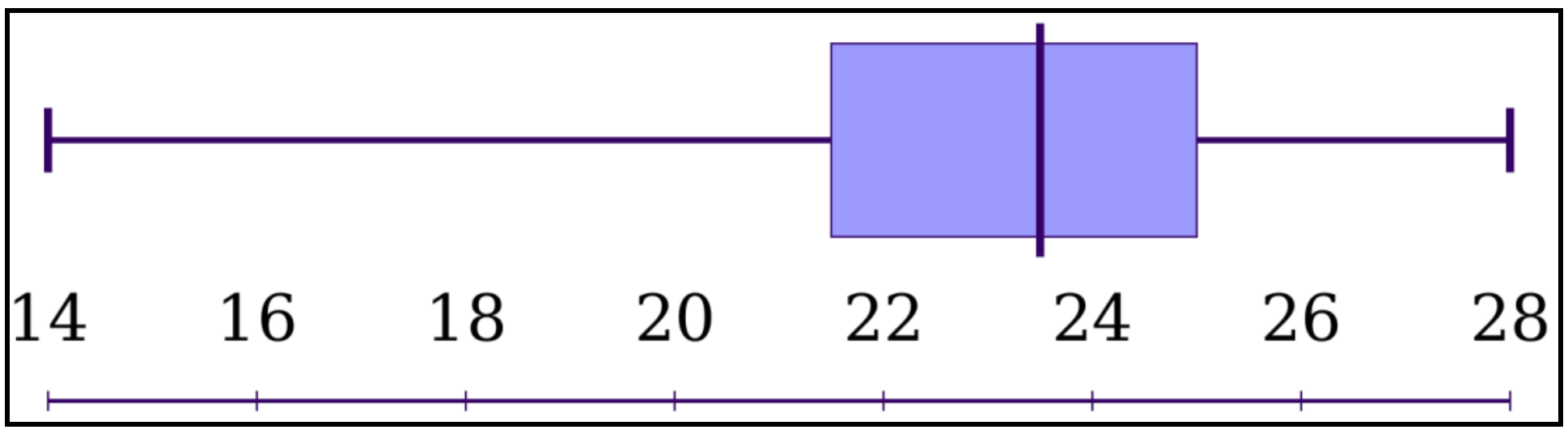

*Figure 9. Control Group marks out of 30 (Continuous Evaluation) in DevOps Course*

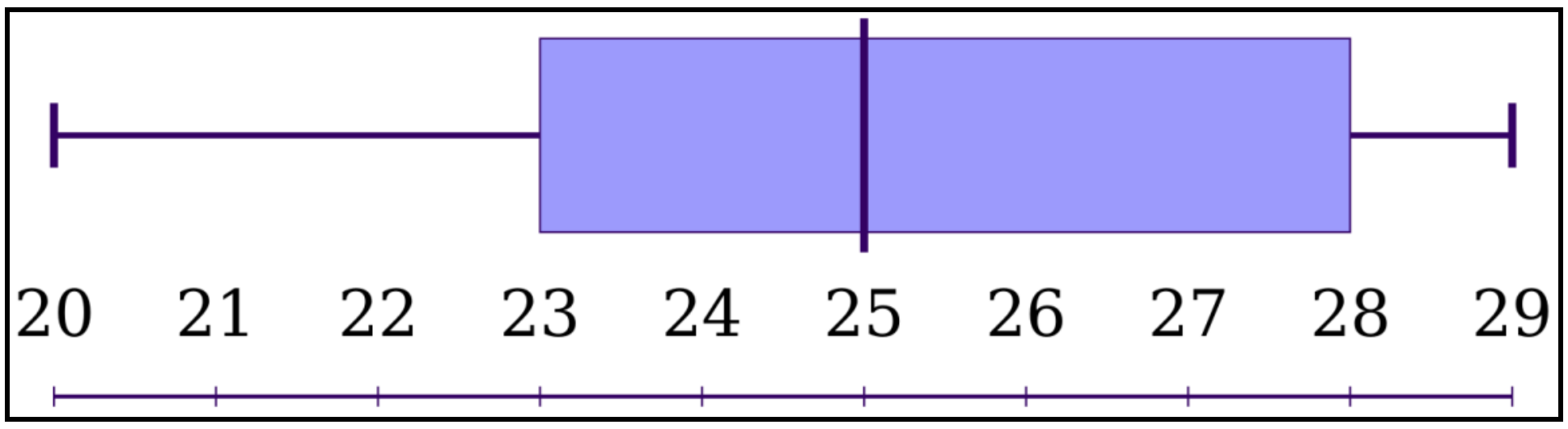

*Figure 10. Experimental Group marks out of 30 (Continuous Evaluation) in DevOps Course*

The Box-Whisker plot in Figure 9 & Figure 10 shows that the experimental group has more marks than the control group in continuous evaluation in each of the quartiles of the experimental group. The distribution of the marks for the experimental group in continuous evaluations is skewed towards the right side. This indicates that the students who participated in the Competition have more marks in continuous evaluations than those who did not participate in the Competition.

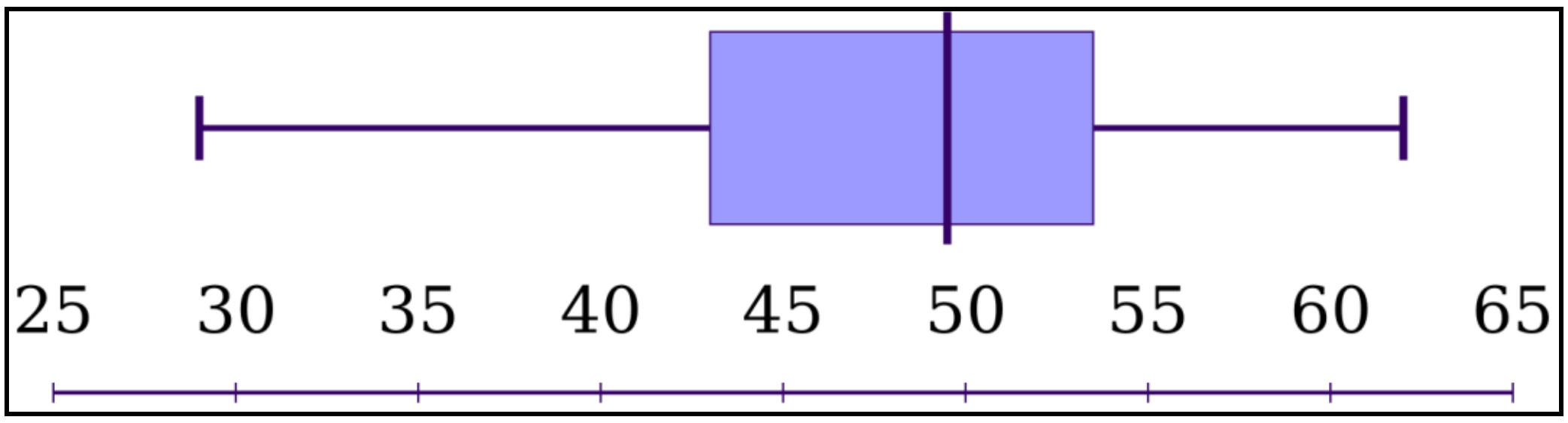

*Figure 11.Control Group marks out of 70 (End-of-Semester Examination) in DevOps Course*

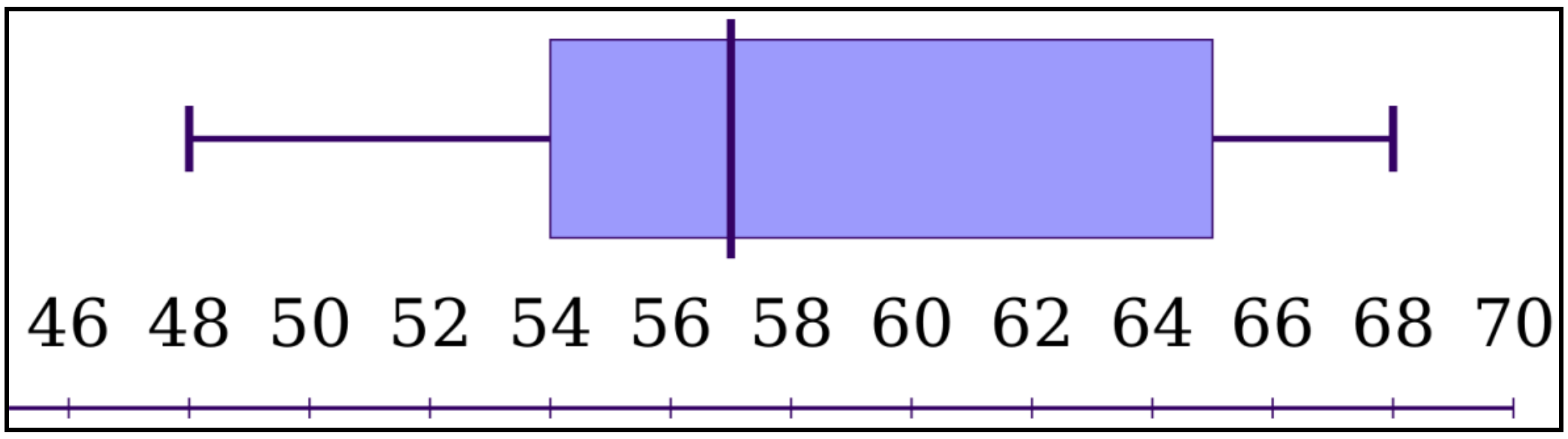

*Figure 12. Experimental Group marks out of 70 (End-of-Semester Examination) in DevOps Course*

The Box-Whisker plot in Figure 11 & Figure 12 shows that the experimental group has more marks than the control group in end-of-semester examinations in each quartile of the experimental group. The distribution of the marks out of 70 for the experimental group is skewed towards the right side. The students who participated in the Competition have more marks in end-of-semester evaluations than those who did not participate in the Competition. The examination department masks the answer sheets of the end-of-semester examination before evaluations by the examiner. Hence there is no possibility of bias.

Following the Normal distribution rules, the Marks obtained by the Student less than Q1-1.5*IQR and marks obtained greater than Q3+1.5*IQR are the outliers. Therefore, the outliers are missing in the given data set. The A/B Test shows that the hypothesis is proven as the experimental group has secured more marks and won the prizes.

*Table 2*

**Statistics of the Marks out of 100 Obtained by the Experimental Group**

| | | | | |
|---|---|---|---|---|
| Experimental GroupSize | 15 | | Maximum Marks | 96 |
| Median(Second Quartile Q2) | 81 | | First quartile (Q1) | 77 |
| Minimum Marks | 72 | | Third quartile (Q3) | 93 |
| Inter-Quartile Range (IQR) | 16 | | Lower End Outliers | None |
| Mean μ | 83.93333 | | Upper End Outliers | None |
| Standard Deviation ς | 8.655854 | | μ - ς | 75.27748 |
| | | | μ + ς | 92.58919 |

**b) PR Submission and Prizes**

As shown in figure 13, there are 64 PRs as the sum of a minimum 4 PRs contribution, and an additional 26 PRs show that the PG students have actively participated in the activity. Each PR submission consists of related activities, such as writing a program for the specific functional requirement in the specific programming language and submission using Git and GitHub to the repository. For example, one PR submission requires the knowledge of the Git Commands and the Github usage skills.

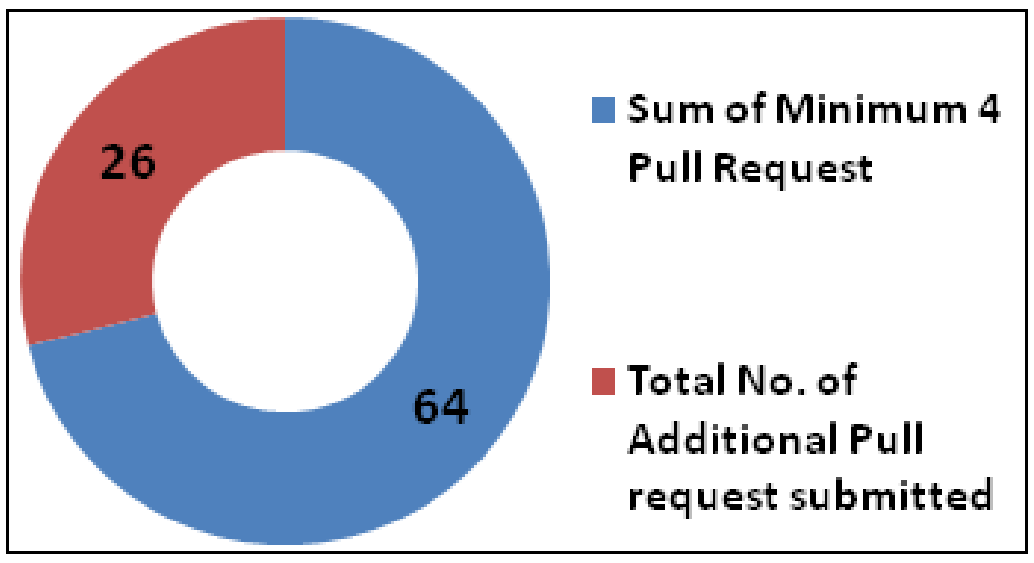

*Figure 13. Hacktoberfest Github Pull Request Statistics*

The outcome of the Course is in terms of the prizes obtained by the students after participation in the Hacktoberfest competition. Figure 14 shows the T-Shirts and stickers as prizes won by the participant students from the experimental group.

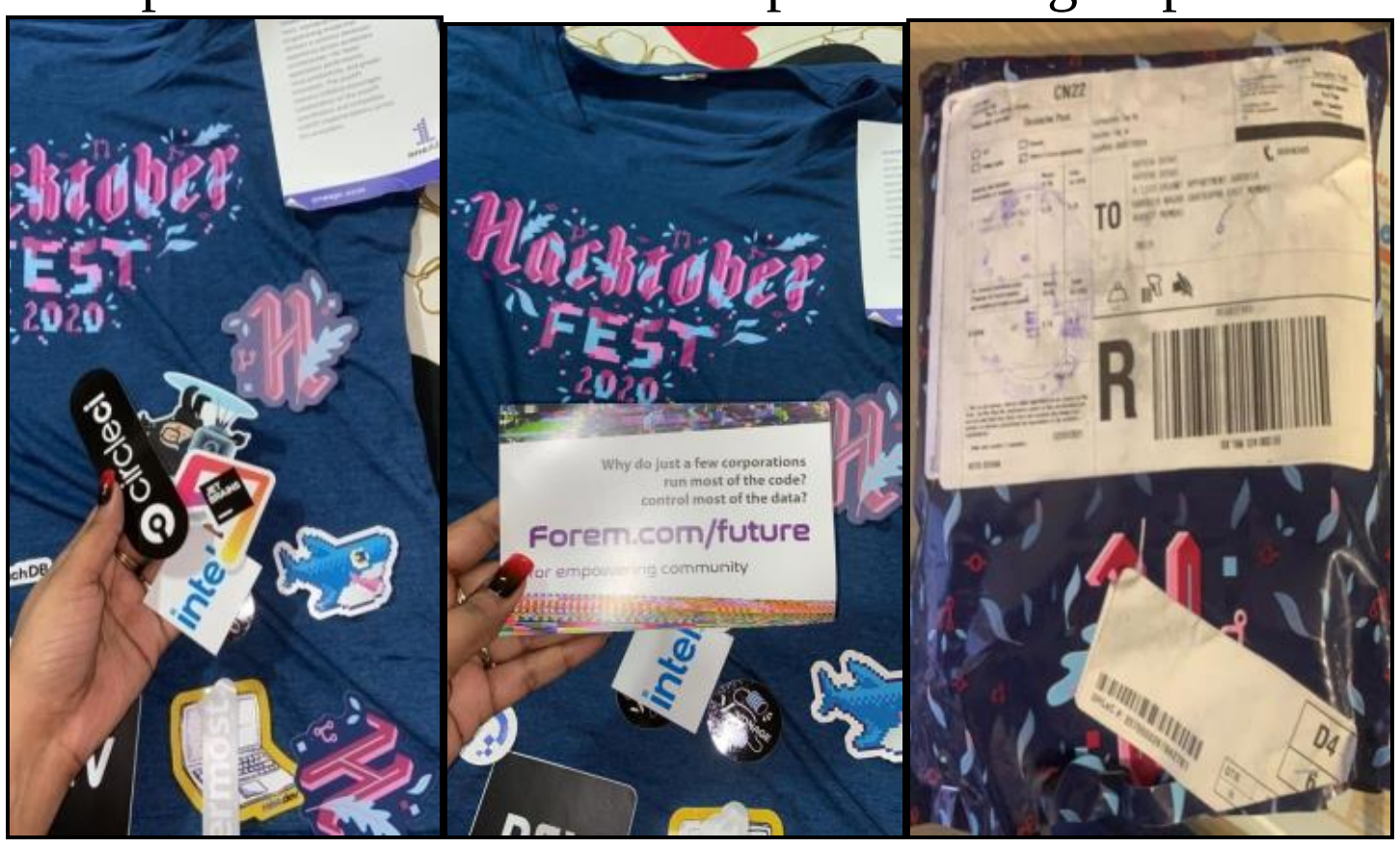

*Figure 14.- Student Rewards after attending the Hacktoberfest Competition*

**5. CONCLUSIONS AND PROSPECTS FOR FURTHER RESEARCH**

The reward-based encouragement, participatory teaching, and motivation to the students are vital factors. In the current IT industry scenario, DevOps has many tools to be covered in the curriculum. The organizers of the competition award the prizes in the form of T-Shirts. All the 16 participants of Hacktoberfest, including the 13 PG students, 02 UG students, and one faculty member, won the Prizes.

The sample was not explicitly divided into the two groups. However, after the students' participation in the Competition, the experimental and control groups were naturally created. There were no explicit marks assigned for participation in the Competition. Interestingly, it is observed that the experimental group who participated in the Competition has performed significantly better than the control group. Using the concept of Normal Distribution, $\mu \pm 2\varsigma$ has not shown any outlier in the marks obtained by the experimental group. Here the $\mu$ is mean, and $\varsigma$ is the Standard Deviation. The outlier is also missing when tested using the Q1±1.5*IQR. Here the Q1 is the First Quartile, and IQR is the Inter Quartile Range. The tangible benefits in terms of outcomes of the research activity were visible. The experimental group has performed better in terms of the marks obtained compared to the control group.

The research shows that 81.25 % of the PG and 12.5 % of UG students' participated in the indirect Gamification. The overall PG participation is 76.47% of the entire PG class. The UG participation is 4% of the entire UG class. The students have experienced peer learning. The indirect Gamification has worked well for effective learning outcomes of the highly technical DevOps Course. The intangible outcomes were also equally observed as the concepts learned by the participants were practically used in the Git and Github tools. Students have clearly understood how to contribute to the repositories maintained by the other owners. Moreover, students have also learned that they can maintain repositories, and after tagging, they can participate in the Competition.

The involvement of the two UG students has played an important role. When UG students perform well in the PR submission, it creates performance excitement among the PG students. On the other side, the motivation by the Faculty through the online class conduction, Whatsapp communication, and telephonic communication encouraged the students. The contribution of each factor is unmeasured; however, the outcomes are visible in terms of the performance grading obtained by the students.

In future work, the effect of learning abilities, age group, learning motivation factors will be investigated. The LMS Moodle is already in use, which has direct Gamification like providing batches, grades, and stars to the participant students.

## REFERENCES (TRANSLATED AND TRANSLITERATED)

# ІНТЕГРАЦІЯ НЕПРЯМОЇ ГЕЙМІФІКАЦІЇ В НАВЧАЛЬНИЙ КУРС З РОЗРОБКИ ТА ЕКСПЛУАТАЦІЇ (DEVOPS)

**Манодж Деваре**
доктор філософських наук, професор, директор Інституту інформаційних технологій Аміті, Університет Аміті Махараштра, Мумбаї, Індія
ORCID ID 0000-0002-9530-3914
*mhdevare@mum.amity.edu*

**Анотація.** У роботі досліджується вплив непрямої гейміфікації на навчання курсу «Розробка та експлуатація» (DevOps). Команда розробників програмного забезпечення, що складається з кількох учасників, координує свої дії в спільному робочому середовищі для досягнення конкретних заздалегідь визначених цілей шляхом регуляції та контролю. У курсі DevOps встановлення та налаштування різних програмних інструментів використовуються в контексті розробки програм. Завдання тренера – навчити використовувати інструменти DevOps для досягнення поставлених навчальних цілей. Створене освітнє середовище повинно мотивувати учнів до навчання. Педагогічні методи та ІКТ відіграють важливу роль у процесі передачі знань. Для досягнення поставленої мети навчання необхідна мотивація та підтримка студента під час проходження курсу. Автор використав незалежний конкурс як технологію непрямої гейміфікації з метою досягнення результатів навчального курсу. Студентам було запропоновано взяти участь у конкурсі Hacktoberfest з метою використання своїх практичних навичок, отриманих під час проходження курсу. Студенти бакалаврату (UG – Under Graduate) та аспірантури (PG – Post Graduate) пройшли регулярні сесії DevOps. Обидві групи студентів спілкувались на одній платформі. Студенти, які швидше внесли дані PR до GitHub, поділилися своїм досвідом з іншими учасниками. Кількість студентів PG (68,75%) значно перевищила учасників-студентів UG. Основним критерієм виконання поставленого завдання було мінімум чотириразове внесення даних PR у GitHub та збереження їх у репозитарії. Активна участь невеликої кількості студентів UG стала мотиваційним фактором для студентів PG. Розподіл Гауса за оцінками, отриманими експериментальною групою, показує відсутність викидів. Дослідження показує, що ефективність непрямої гейміфікації залежить від вікової групи, рівня, змісту курсу та середовища навчання. Участь викладача в конкурсі під час навчальної діяльності підвищує бажання студента виконати завдання. Експериментальна група з 15 студентів перевершила за отриманими оцінками контрольну групу з 52 студентів.